\documentclass[letterpaper, 10 pt, conference]{ieeeconf}  

\IEEEoverridecommandlockouts                              

\usepackage{graphicx} 
\usepackage{amsmath} 
\usepackage{algorithm}
\usepackage{algpseudocode}
\usepackage{tabularx}
\usepackage{booktabs}
\usepackage{tfrupee}
\usepackage{inputenc}
\usepackage{booktabs}
\usepackage{multirow}
\usepackage{graphicx}
\usepackage{soul}
\usepackage{xcolor}
\usepackage{url}
\usepackage{subcaption}
\title{\LARGE \bf
Predictive Tree-based Virtual Keyboard for Improved Gaze Typing}

\author{Hrushikesh Etikikota$^{1}$, Yogesh Kumar Meena$^{*2}$
\thanks{$^{1,2}$ Hrushikesh Etikikota and Yogesh Kumar Meena are with Human-AI Interaction (HAIx) Lab, IIT Gandhinagar, India
        {\tt\small yk.meena@iitgn.ac.in}}%
}

\begin{document}

\maketitle
\thispagestyle{empty}
\pagestyle{empty}

\begin{abstract}

On-screen keyboard eye-typing systems are limited due to the lack of predictive text and user-centred approaches, resulting in low text entry rates and frequent recalibration. This work proposes integrating the prediction by partial matching (PPM) technique into a tree-based virtual keyboard. We developed the Flex-Tree on-screen keyboard using a two-stage tree-based character selection system with ten commands, testing it with three degree of PPM (PPM1, PPM2, PPM3). Flex-Tree provides access to 72 English characters, including upper- and lower-case letters, numbers, and special characters, and offers functionalities like the delete command for corrections. The system was evaluated with sixteen healthy volunteers using two specially designed typing tasks, including the hand-picked and random-picked sentences. The spelling task was performed using two input modalities: (i) a mouse and (ii) a portable eye-tracker. Two experiments were conducted, encompassing 24 different conditions. The typing performance of Flex-Tree was compared with that of a tree-based virtual keyboard with an alphabetic arrangement (NoPPM) and the Dasher on-screen keyboard for new users. Flex-Tree with PPM3 outperformed the other keyboards, achieving average text entry speeds of 27.7 letters/min with a mouse and 16.3 letters/min with an eye-tracker. Using the eye-tracker, the information transfer rates at the command and letter levels were 108.4 bits/min and 100.7 bits/min, respectively. Flex-Tree, across all three degree of PPM, received high ratings on the system usability scale and low-weighted ratings on the NASA Task Load Index for both input modalities, highlighting its user-centred design.

\end{abstract}


\section{Introduction}

Globally, over 12.2 million new people suffer from strokes each year, often resulting in the loss of motor and speech abilities~\cite{WSOReprt}. This creates significant communication challenges for those affected. Developing assistive devices to aid communication for disabled individuals aligns with major global development goals~\cite{howden2017sdg}. Previous research in human-computer interaction has focused on enhancing communication for differently-abled users~\cite{ascari2021computer,pinheiro2011alternative, meena2018toward}.

Technological advancements have introduced various on-screen keyboards (i.e., virtual keyboards) like clickers, PenFriend, and Dasher~\cite{OSK, ward2000dasher} for typing and spelling tasks. EEG-EOG-based virtual keyboards~\cite{hosni2019eeg} offer potential assistance for differently-abled individuals. Still, they face challenges such as high setup time and low information transfer rates (ITR)~\cite{hosni2019eeg}. Webcam-based virtual keyboard \cite{chakraborty2019eye} detects eye gaze and blinking without needing wearable devices. Yet, poor gaze detection, noisy environments, and involuntary blinking can limit their effectiveness.

Gaze control, less affected by motor and speech disabilities, presents a promising alternative~\cite{cecotti2018multimodal}. Eye-tracking-based virtual keyboards could be a viable solution for these users. Various gaze-based virtual keyboards have been developed with different mechanisms, such as using eye gaze for command detection and selection~\cite{meena2019design}. In contrast, others combine gaze with technologies like soft switches or mice for faster communication~\cite{meena2019design}. Two-level tree-based interfaces have been suggested for languages with many characters~\cite{cecotti2019multiscript}, and a novel method to optimize the position of displayed items in gaze-controlled tree-based menus has been proposed~\cite{meena2018toward}, considering frequency and alphabetic order. However, these interfaces still need predictive text and user-centred approaches to improve typing speed.

Unlike two-level tree-based interfaces, the Dasher on-screen keyboard~\cite{ward2000dasher} employs a continuous text entry method where characters are selected and deleted based on the user's eye-gazing position on the screen. While this approach offers a seamless typing experience, it can be challenging for new users, especially when using eye-tracking for languages with large character sets~\cite{tuisku2008now}. The Dasher method often suffers from low initial transfer rates and slower adaptability, making it less intuitive for beginner users. In eye-tracking based interfaces, two primary selection methods are used: dwell-based and dwell-free. Dwell-based methods require users to fixate on a target item, usually a character, to make a selection. In contrast, dwell-free methods utilize alternative input techniques, such as soft switches or sEMG-based hand gestures, to perform selections without prolonged fixation~\cite{meena2019design}. Both methods have their advantages and challenges, depending on the user's familiarity and the complexity of the language.

To overcome challenges like slower adaptability and low information transfer rates in on-screen keyboards, it is essential to develop designs that incorporate predictive text and user-centered approaches. This study aims to enhance gaze typing by proposing predictive text integration and on-screen keyboard improvements. Our contributions include: 1) integrating the prediction by partial matching (PPM) algorithm into tree-based virtual keyboards to predict the next character, and 2) benchmarking the performance of PPM-based and non-PPM-based on-screen keyboards, including one-level and two-level tree-based interfaces, with mouse and eye-tracker access



The paper is organized as follows: Section II introduces the method for integrating the Prediction by Partial Matching (PPM) algorithm. Section III outlines the system overview of the Flex-Tree and Dasher on-screen keyboards. Section IV details the experimental protocol, including design and operation. Section V presents the results of experiments 1 and 2. Section VI covers the subjective evaluation, followed by the discussion and conclusion in Section VII.


\section{Proposed prediction by partial matching for tree-based interfaces}


We integrated the Prediction by Partial Matching (PPM) technique~\cite{drinic2003ppm} with tree-based virtual keyboards~\cite{cecotti2019multiscript} to enhance typing performance and design a more efficient interface. By implementing the PPM algorithm, we used dictionaries to create a model that predicts the next probable character. This model, combined with the PPM technique in Algorithm~\ref{alg:myalgorithm}, was used to assign the most probable characters to the command buttons in levels 1 and 2 of the Flex-Tree on-screen keyboard layout. 

The \texttt{PredModel} is built using a dictionary-based implementation of PPM~\cite{hu2010improving}. It is a nested dictionary where the outer keys represent the context (the last K characters of user-typed text), the inner keys represent the next probable characters, and the inner values indicate the frequency of those characters. For example, given the text 'Hello' and using PPM2 (where 2 denotes the last two characters of user-typed text i.e., context), the PredModel provides the following suggestions: PredModel = \{``He": \{`l': 1\},``el": \{`l': 1\}, ``ll``: \{`o': 1\}, ``lo": \{`\$': \}\}. Here \$ denotes the end of the string, with the keys representing the context and the values showing the predicted next characters.



\begin{figure*}[]
    \centering
    \begin{subfigure}[b]{0.38\textwidth}
        \centering
        \includegraphics[width=\linewidth]{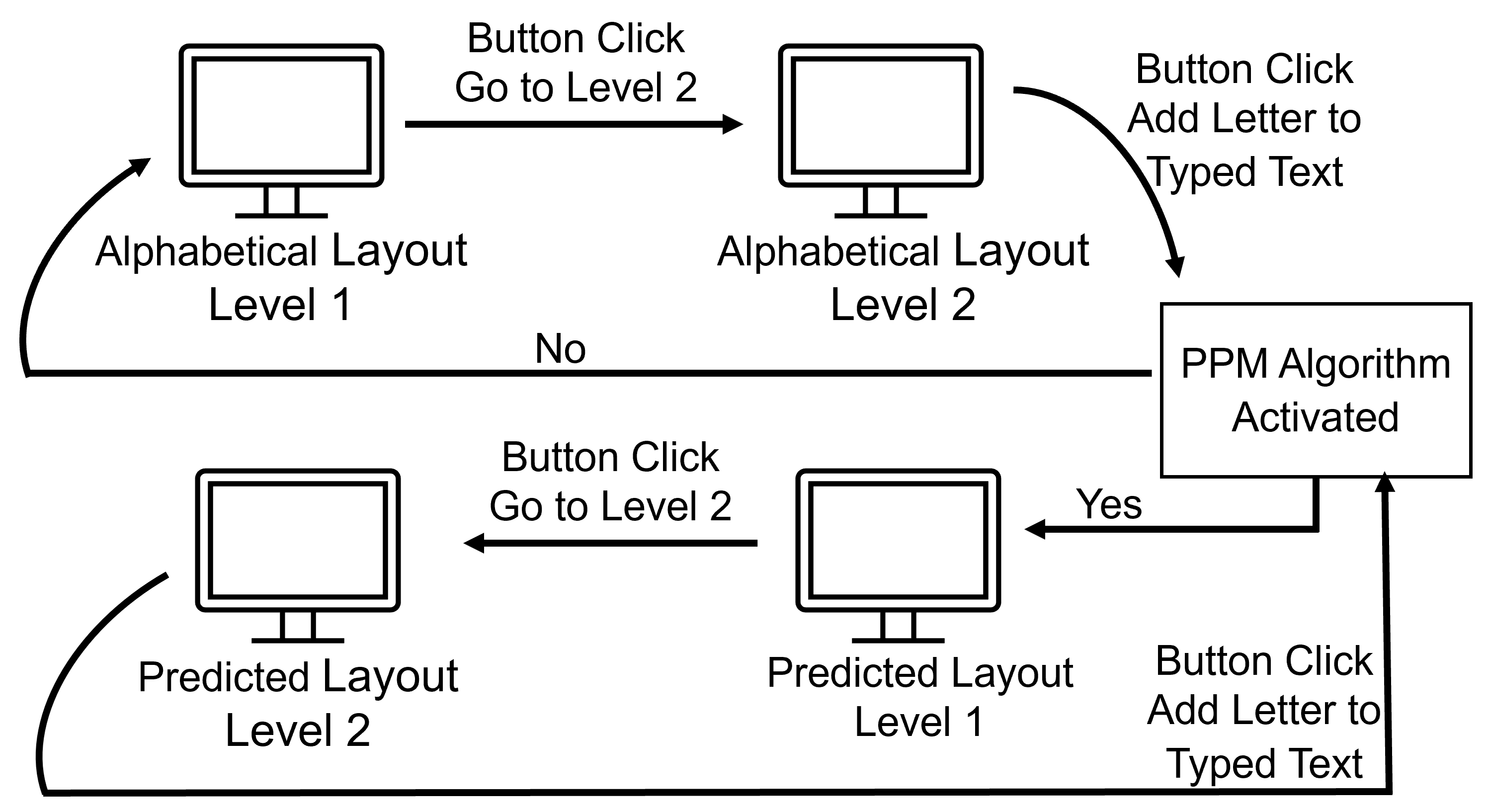}
        \label{fig:sub2}
    \end{subfigure}
    \hfill
    \begin{subfigure}[b]{0.35\textwidth}
        \centering
        \includegraphics[width=\linewidth]{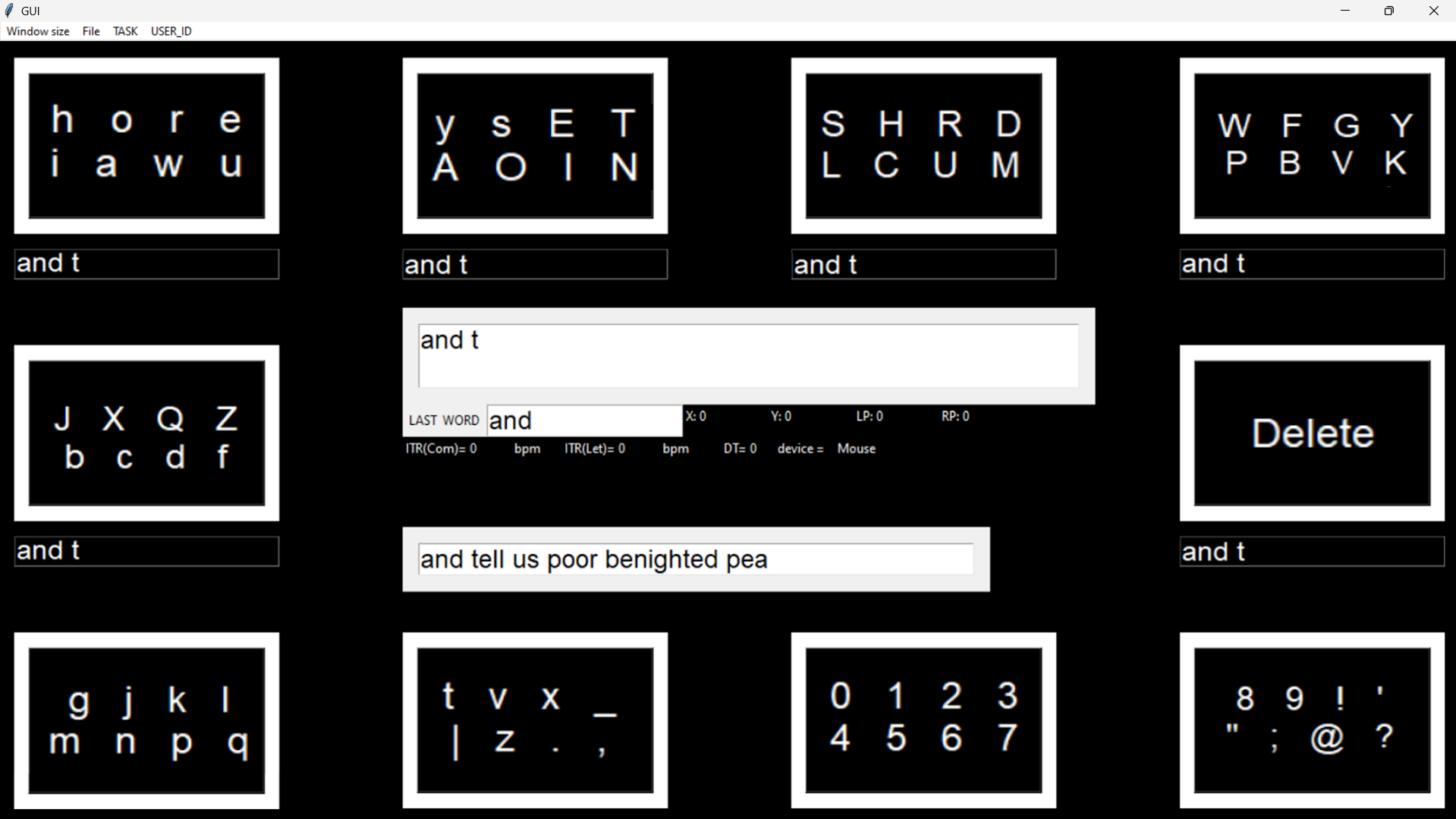}
        \label{fig:sub3}
    \end{subfigure}
    \hfill
    \begin{subfigure}[b]{0.24\textwidth}
        \centering
        \includegraphics[width=\linewidth]{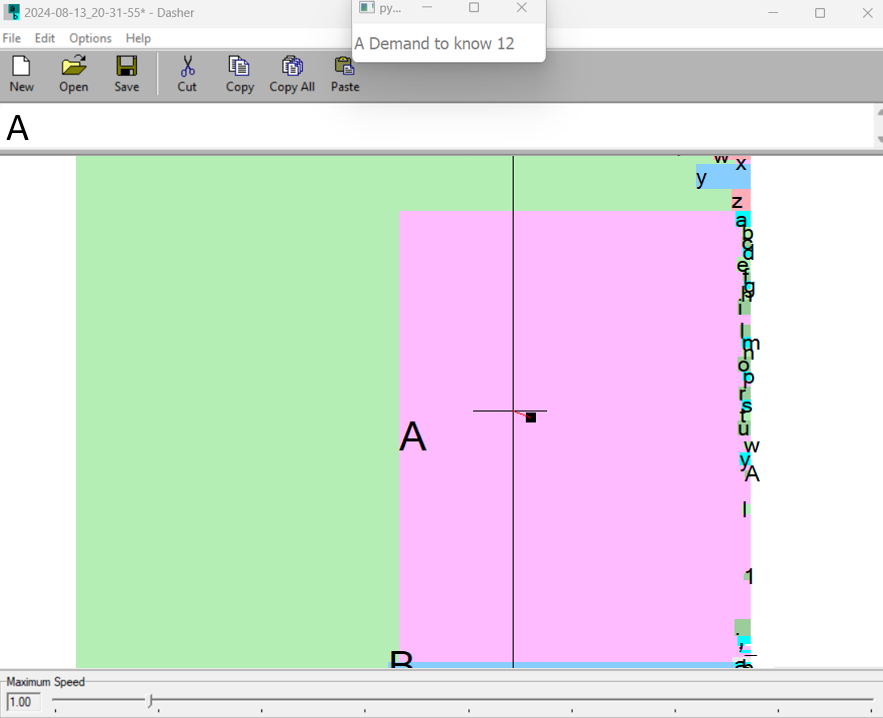}
        \label{fig:sub1}
    \end{subfigure}
    \caption{Left: Workflow diagram of Flex-Tree on-screen keyboard with proposed PPM method. Middle: GUI of the proposed Flex-Tree with PPM2 in level one after typing ``and t” is shown in the output text display. Right: GUI of the reproduced Dasher on-screen keyboard includes the input text display for typing spelling task~\cite{ward2000dasher,tuisku2008now}. }
    \label{fig:main}
\end{figure*}

Integrating PPM into the tree-based interface (see in Fig.~\ref{fig:main} (Left)) involves obtaining the most probable characters for a given context using \textit{PredModel}, ensuring smooth transitions between levels, arranging characters on the screen, handling new contexts outside \textit{PredModel}, and enabling easy navigation. Algorithm~\ref{alg:myalgorithm} was designed to address these conditions and ensure seamless integration. The algorithm uses inputs \texttt{Change\_level\_to} takes two values 1 or 2 indicate which level to go next, $K$ denote the degree of PPM, and \texttt{Command\_id} denotes the ID [1 to 10] of the command the user clicked either in  level 1 or  level 2.


Transitioning from level 1 to level 2 is straightforward, \texttt{Change\_level\_to} is set to 2, and current\_text becomes the text on button\_id. The candidates list includes the first 4 characters of current\_text, along with ``GO BACK," ``DELETE," and the last 4 characters of current \_text. Moving from level 2 to level 1 adds the character associated with button\_id to Text\_ Entered and returns to level 1. If Text\_ Entered is shorter than $K$, all characters are grouped alphabetically into sets of 8. Otherwise, buttons display strings of 8 characters in this order: the most probable next characters from \textit{PredModel}, the most frequent unused characters, and finally, the remaining characters in the language.

These steps generate nine strings of length 8, which are added to the Candidates list. This list contains the text for each command button in level 1. We place ``DELETE" at the sixth position in Candidates, as it is consistently positioned there at both levels. The Candidates list now has a length of 10, with indices from 1 to 10 representing the text for buttons 1 through 10 in level 1.

\begin{algorithm}[]
 \small

 \noindent\hrule
 \begin{algorithmic}[1]
  \State \textbf{Input:$Change\_level\_to$}-Indicates the next level, either 1 or 2.
  \State \textbf{Input:$K$}-Degree of PPM used  or length of context 
  \State \textbf{Input:$command\_id$}- ID of Button just clicked in either level
  \State \textbf{Output:$Candidates$}- A list of strings representing the text in each button command

  \If{$\text{Change\_level\_to} = 1$}
  
   Add the character on command button to the \text{text\_entered}
    \If{$\text{text\_entered} < k$}
      \State candidates $\gets$ group all the characters alphabetically 
    \Else
      \State sets1 $\gets$ Most Probable Characters are grouped.
      \State sets2 $\gets$ Most frequent letters in language are grouped
      \State sets3 $\gets$ Remaining characters are grouped
      \State add set1, set2, and set3 to candidates
    \EndIf
  \Else
    \State set1 $\gets$ 8 characters from the $command\_id$ in level 1
    \State add set1 to candidates
    \State add 'BACK' at fifth position of the candidates
  \EndIf
  \State add 'DELETE' at sixth position of the candidates
  \State \textbf{return Candidates} 
 \end{algorithmic}
Note: Grouping in the algorithm always indicates the formation of strings of length 8 characters.
 \caption{PPM for Tree-Based Interfaces}
 \label{alg:myalgorithm}
\end{algorithm}

\section {System Overview}
\subsection {Flex-Tree on-screen keyboard}

The developed graphical user interface (GUI) of Flex-Tree mainly consists of an input text display, a total of ten commands, an output text display, and a text display showing the last five characters designed below each input command.  This design allows users to view previously typed characters without shifting their gaze to the central output display. The layout and design are illustrated in Fig.~\ref{fig:main} (Middle).










Flex-Tree has two levels for typing 72 characters, including upper and lowercase English letters, numbers, and select special characters (.,"';?\textbar{}\_). Level 1 contains 10 commands, each representing 8 characters, except for command 6, which is always ``DELETE" for erasing the last typed character. Level 2 also has 10 commands: commands 1-4 and 7-10 each represent a character, command 6 is ``DELETE," and command 5 is ``GO BACK'' allowing a return to level 1 without typing in level 2. Character selection occurs on level 2. First, the user selects a command button (except DELETE), narrowing the selection to 8 characters and moving to level 2. After selecting a command (except DELETE or GO BACK), the chosen character is added to the output text display, and the system returns to level 1.

The order or degree ($K$) of PPM is important in transitioning from alphabetical~\cite{meena2018toward} to predicted layouts. Predicted layouts dynamically adjust the text on command buttons based on user input and the specific PPM algorithm used for the Flex-Tree layout interfaces and their interrelations (see Fig.~\ref{fig:main} (Left)). When \textit{K = 0}, Flex-Tree consistently uses alphabetical layouts as described in workflow diagram and illustrated in Fig.~\ref{fig:main} (Left). For $K = t$, Flex-Tree uses alphabetical layouts until the first $t$ characters are typed, then switches to predicted layouts. The text on command buttons in predicted layouts is determined by the candidates generated in workflow diagram, as shown in Fig.~\ref{fig:main} (Left). 


We developed and integrated auditory and visual feedback into Flex-Tree for each command. Auditory feedback announces the sound of typed characters, while visual feedback changes the button border color from dark green to light green as the user gazes at it. This visual cue enhances command selection efficiency, which is important for eye-tracking-based virtual keyboard applications~\cite{meena2018toward,10.14236/ewic/HCI2018.148}.


\subsection {Dasher on-screen keyboard}

The Dasher on-screen keyboard~\cite{ward2000dasher,tuisku2008now} offers a unique text entry interface controlled by continuous pointing gestures. Characters start alphabetically on the right side of the screen. Users point toward desired characters, triggering a zoom effect that centers and enters the selected character into a text box. Dasher’s language model then predicts likely subsequent characters, which dynamically expand around the chosen one for faster selection. To correct entries, users point left to zoom out, returning characters to their original positions. The central vertical line serves as a reference point, pausing actions when the cursor is centered. 

In this work, we reproduced the Dasher on-screen keyboard using the open-source version 4.0.2, setting the speed to 1, the maximum speed achieved by first-time users in experiments by~\cite{tuisku2008now}. We added an overlay input box to display the typing task at the top of the Dasher screen. Fig.~\ref{fig:main} (Right) shows the Dasher GUI with the input box and the interface after typing ``A".



\begin{table*}[]
\centering
\caption{Typing spelling performance using Flex-Tree on-screen keyboard with hand-picked (HandFT) and randomly-picked (RandFT) sentences for the mouse and the eye-tracker modalities. *Speed in letters/min and ITR in bits/min.}
\label{tab:my-table1}
\begin{tabular}{@{}cc|cccccc|cccccc@{}}
\toprule
\multicolumn{2}{c|}{\multirow{3}{*}{Condition}} & \multicolumn{6}{c|}{Mouse} & \multicolumn{6}{c}{Eye-tracker} \\ \cmidrule(l){3-14} 
\multicolumn{2}{c|}{} & \multicolumn{3}{c}{HandFT} & \multicolumn{3}{c|}{RandFT} & \multicolumn{3}{c}{HandFT} & \multicolumn{3}{c}{RandFT} \\ \cmidrule(l){3-14} 
\multicolumn{2}{c|}{} & Speed & \begin{tabular}[c]{@{}c@{}}ITR\_\\ \textit{com}\end{tabular} & \begin{tabular}[c]{@{}c@{}}ITR\_\\ \textit{letter}\end{tabular} & Speed & \begin{tabular}[c]{@{}c@{}}ITR\_\\ \textit{com}\end{tabular} & \begin{tabular}[c]{@{}c@{}}ITR\_\\ \textit{letter}\end{tabular} & Speed & \begin{tabular}[c]{@{}c@{}}ITR\_\\ \textit{com}\end{tabular} & \begin{tabular}[c]{@{}c@{}}ITR\_\\ \textit{letter}\end{tabular} & Speed & \begin{tabular}[c]{@{}c@{}}ITR\_\\ \textit{com}\end{tabular} & \begin{tabular}[c]{@{}c@{}}ITR\_\\ \textit{letter}\end{tabular} \\ \midrule
\multirow{2}{*}{NoPPM} & Mean & 21.7 & 143.9 & 133.7 & 24.1 & 160.5 & 149.1 & 13.3 & 88.4 & 82.1 & 13.8 & 91.8 & 85.3 \\
 & Std & 07.3 & 48.4 & 44.9 & 04.4 & 29.5 & 27.4 & 02.9 & 19.5 & 18.1 & 02.1 & 13.9 & 12.9 \\
\multirow{2}{*}{PPM1} & Mean & 23.7 & 157.4 & 146.3 & 26.4 & 175.5 & 163 & 15.3 & 102.1 & 94.9 & 15.2 & 101.1 & 94.0 \\
 & Std & 04.9 & 32.9 & 30.6 & 06.8 & 45.1 & 41.9 & 02.6 & 17.6 & 16.4 & 01.8 & 12.4 & 11.5 \\
\multirow{2}{*}{PPM2} & Mean & 30.3 & 201.6 & 187.3 & 28.3 & 188.1 & 174.8 & 15.7 & 104.6 & 97.2 & 15.6 & 103.5 & 96.2 \\
 & Std & 09.1 & 60.9 & 56.6 & 07.4 & 49.1 & 45.7 & 03.4 & 22.6 & 21.0 & 02.8 & 18.8 & 17.4 \\
\multirow{2}{*}{PPM3} & Mean & 27.5 & 182.9 & 169.9 & 27.7 & 184.3 & 171.2 & 16.0 & 106.5 & 99.0 & 16.3 & 108.4 & 100.7 \\
 & Std & 07.7 & 51.3 & 47.7 & 08.9 & 59.4 & 55.2 & 02.6 & 17.8 & 16.5 & 02.9 & 19.7 & 18.3 \\ \bottomrule
\end{tabular}
\end{table*}

\begin{table}[]
\caption{Typing spelling performance with Dasher on-screen keyboard with hand-picked (HandDS) and randomly-
picked (RandDS) sentences where speed in letters/min and ITR in bits/min. Comparison of deletion time with with Flex-
Tree and Dasher in seconds/letter.}
\label{tab:tab3}
\begin{tabular}{@{}cc|cccc|cc@{}}
\toprule
\multicolumn{2}{c|}{\multirow{3}{*}{Condition}} & \multicolumn{4}{c|}{Dasher Typing Performance} & \multicolumn{2}{c}{Deletion Time} \\ \cmidrule(l){3-8} 
\multicolumn{2}{c|}{} & \multicolumn{2}{c|}{Mouse} & \multicolumn{2}{c|}{Eye-tracker} & \multicolumn{2}{c}{Mouse} \\ \cmidrule(l){3-8} 
\multicolumn{2}{c|}{} & Speed & \multicolumn{1}{c|}{\begin{tabular}[c]{@{}c@{}}ITR\_\\ \textit{letter}\end{tabular}} & Speed & \begin{tabular}[c]{@{}c@{}}ITR\_\\ \textit{letter}\end{tabular} & \begin{tabular}[c]{@{}c@{}}Flex-\\ Tree\end{tabular} & Dasher \\ \midrule
\multirow{2}{*}{HandDS} & Mean & 09.1 & \multicolumn{1}{c|}{56.6} & 06.8 & 42.1 & 01.4 & 04.2 \\
 & Std & 03.0 & \multicolumn{1}{c|}{18.6} & 03.2 & 20.3 & 00.1 & 01.0 \\
\multirow{2}{*}{RandDS} & Mean & 10.3 & \multicolumn{1}{c|}{63.8} & 04.9 & 30.5 & 01.4 & 04.5 \\
 & Std & 03.3 & \multicolumn{1}{c|}{20.3} & 01.7 & 10.5 & 0.16 & 01.0 \\ \bottomrule
\end{tabular}
\end{table}

 \section{Experimental Protocol} 

\subsection{Participants}

A total of sixteen healthy volunteers (2 females) in age range of 22--31 years (25.2$\pm$1.80) participated in this
study. Four participants had vision correction. No participant had prior experience using an eye-tracker or mouse with both on-screen keyboard applications. Participants were advised regarding the purpose and nature of the study. We paid each participant Rs.100 ($\approx$ \$1.21) for their participation. We followed an Institute Ethical Committee (IEC) approved informed consent procedure with participants while conducting the experiments.


\subsection{Design and operational procedure}

The experimental protocol is designed to write four predefined sentences (two sentences for Flex-Tree and two sentences for Dasher). For Flex-Tree, first, a hand-picked sentence (HandFT) is defined as \textit{``A Demand to know what happened"}. HandFT is specially designed so that each command has an equal chance to be selected and overcome the bias. Second, we randomly selected a sentence from training data~\cite{ward2000dasher} and named it as RandFT. RandFT is defined as \textit{``and tell us poor benighted pea"}. However, both RandFT and HandFT consist of 36 characters (72 commands if there are no errors).

From our early experiments, we observed that users found it difficult to write HandFT and RandFT with the Dasher application using the eye-tracking modality as they are very long. Therefore, due to the duration of the experiment, we reduce the size of the typing task (nearly ~50\%) sentences for the Dasher application. For Dasher, the random-picked sentence (RandDS) is defined as \textit{``and tell us poor"}. hand-picked sentence (HandDS) is defined as \textit{``A Demand to know"}. However, both RandSD and HandDS consist of 16 characters (32 commands if there are no errors). 

The experiment employed two distinct input methods: a wired mouse and a portable eye-tracking device (The Eye Tribe)~\cite{meena2019design,meena2016novel}. Users interacted with the virtual keyboard by clicking commands when using the mouse. In the case of eye-tracking, participants calibrated their gaze using the calibration window provided by the eye-tracking device's SDK.
Both input modalities were utilized for all the experimental conditions.

For command selection with the eye-tracker, we used the asynchronous mode algorithms developed by~\cite{meena2019design}. In this mode, users must maintain their gaze on a target for a predetermined duration (i.e., dwell time). For example, with a dwell time of 1.5 seconds, the user must look at the target for 1.5 seconds to select it. We employed a 1.5-second dwell time in our eye-tracking experiments~\cite{meena2018toward}.

To assess overall performance across 24 different conditions, we divided our experiment into two parts: 1) Experiment 1: Evaluating hand-picked and random-picked sentences for both applications using a mouse and eye-tracker, and 2) Experiment 2: Measuring deletion times for a task using a mouse only, applied to both applications. 


\subsection{Experiment 1}

In this experiment, we evaluate the performance of Flex-Tree and Dasher on-screen keyboards with typing tasks both inside and outside the training text corpus. For each hand-picked and random-picked sentence, we conduct three sub-experiments: 1) PPM versus NoPPM with a mouse, 2) PPM versus NoPPM with an eye-tracker, and 3) Dasher with a mouse versus an eye-tracker.


\textit{1. PPM versus NoPPM with mouse:} We tested four conditions on Flex-Tree using a mouse: NoPPM and three degrees of PPM (PPM1, PPM2, and PPM3). Users typed hand-picked (HandFT) and random-picked (RandFT) sentences with each algorithm setting. This allowed us to evaluate how different PPM degrees affect typing speed and establish a baseline for comparison with the eye-tracker


\textit{2. PPM versus NoPPM with eye-tracker:} We tested the same conditions on the Flex-Tree using an eye-tracker. Users typed hand-picked (HandFT) and random-picked (RandFT) sentences with the eye-tracker. Our goal was to evaluate the PPM-based virtual keyboard as an alternative to Dasher for beginner users and assess the impact of PPM on performance, determining its suitability compared to Dasher. 


\textit{3. Dasher with a mouse versus an eye-tracker: } In this experiment, Dasher was tested using both eye-tracker and mouse settings. Users typed hand-picked (HandDS) and random-picked (RandDS) sentences with both input methods. This allows us to compare Dasher’s performance with a mouse versus an eye-tracker and evaluate how beginner users perform with the eye-tracker.


\subsection{Experiment 2}


In the experiments, we assess how deletion actions impact application performance. Quick deletion of mistyped letters indicates a more efficient error correction mechanism. We test deletion times with HandDS and RandDS inputs using a mouse on both Flex-Tree and Dasher on-screen keyboards.


\subsection{Performance evaluation}
We evaluated the proposed GUI layouts using several performance metrics: the number of letters spelled per minute, the information transfer rate (ITR) at both the letter level ($ITR_{letter}$) and command level ($ITR_{com}$)~\cite{ meena2018toward,meena2019design,9964264}, and the mean and standard deviation (mean ± SD) of the time required to produce each command. ITR is calculated from the total number of actions (commands and letters) and the time taken to complete these actions. In our case at command level, the number of possible commands is 10 ($M_{com} = 10$) and letter level is 72 ($M_{letter} = 72$). For ITR estimation, we assumed that all commands and letters are equally likely and that there are no typing errors.

\section{Results}

\begin{figure*}[]
    \centering
    \begin{subfigure}[b]{0.24\textwidth}
        \centering
        \includegraphics[width=\linewidth]{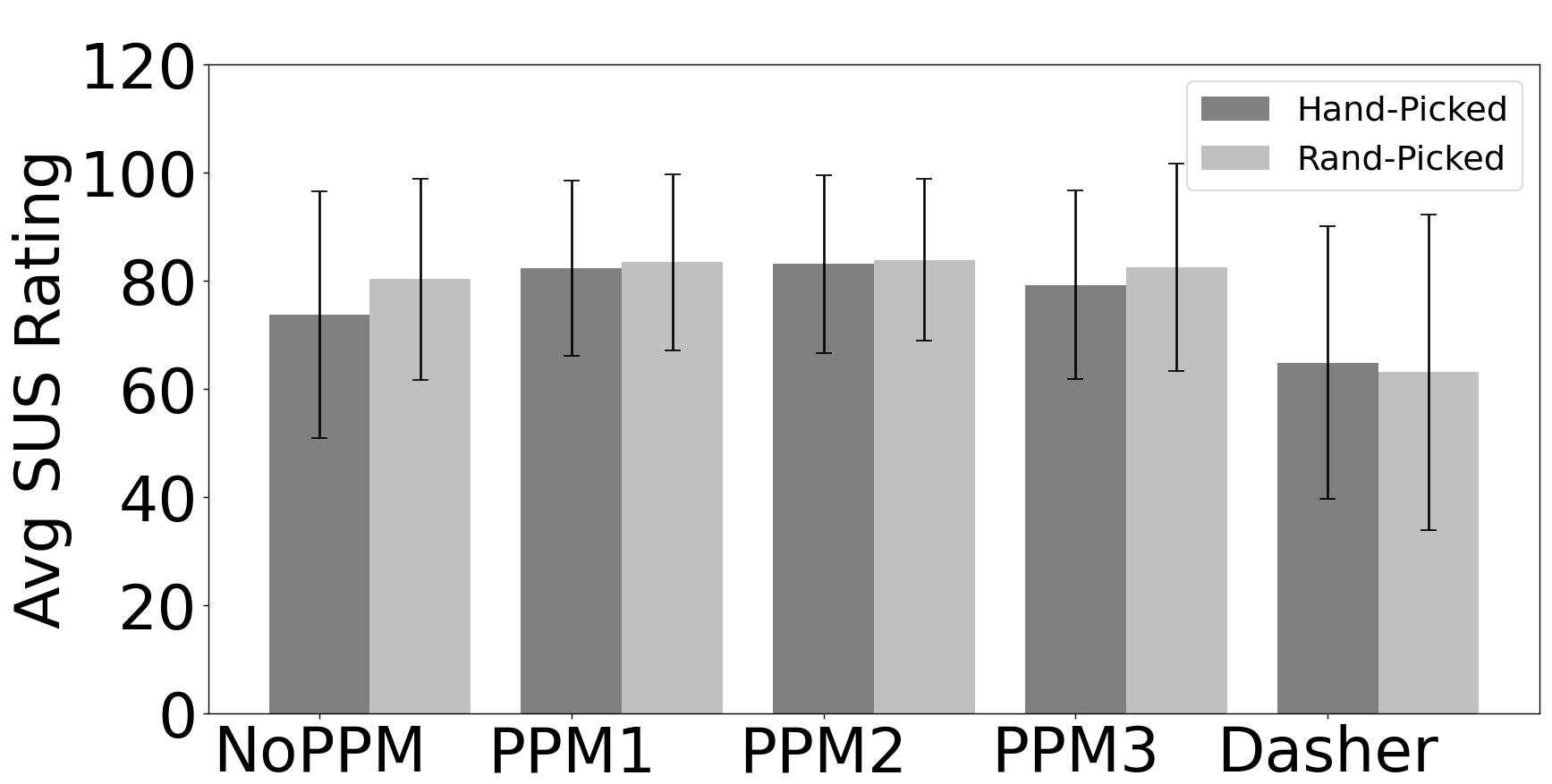}
         \caption{}
        \label{fig:sub2}
    \end{subfigure}
    \begin{subfigure}[b]{0.24\textwidth}
        \centering
        \includegraphics[width=\linewidth]{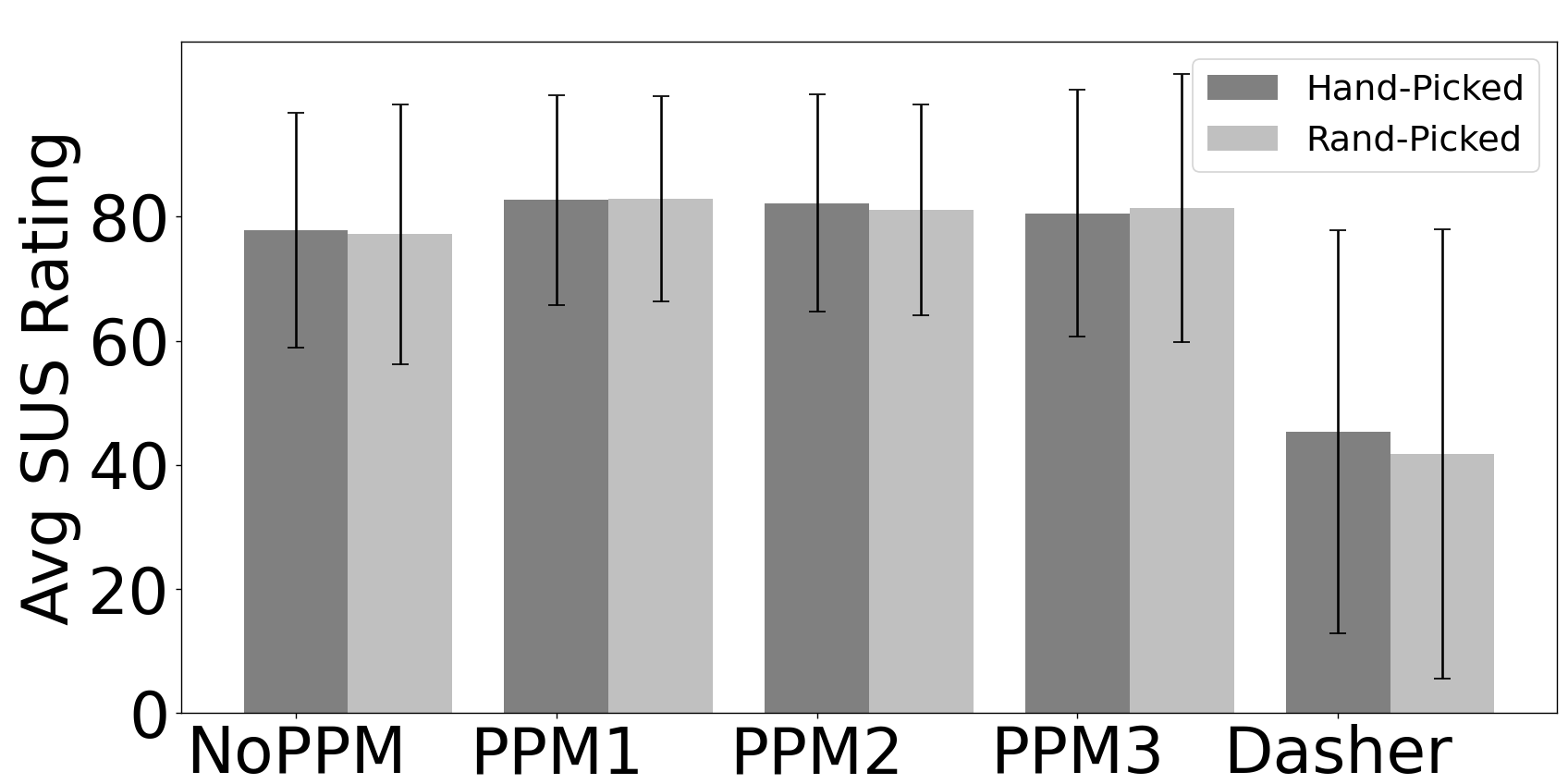}
         \caption{}
        \label{fig:sub3}
    \end{subfigure}
    \begin{subfigure}[b]{0.24\textwidth}
        \centering
        \includegraphics[width=\linewidth]{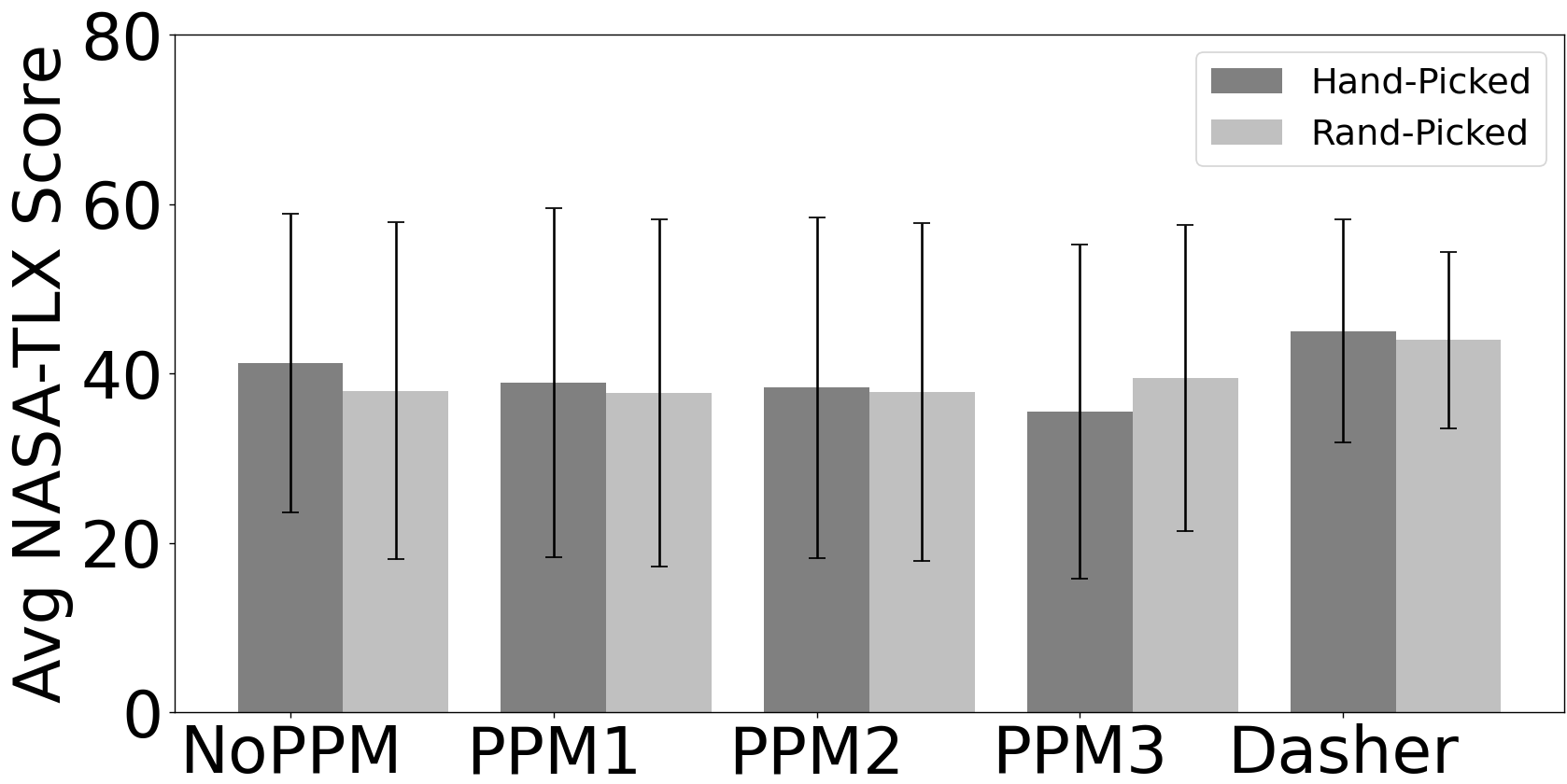}
         \caption{}
        \label{fig:sub1}
    \end{subfigure}
    \begin{subfigure}[b]{0.24\textwidth}
        \centering
        \includegraphics[width=\linewidth]{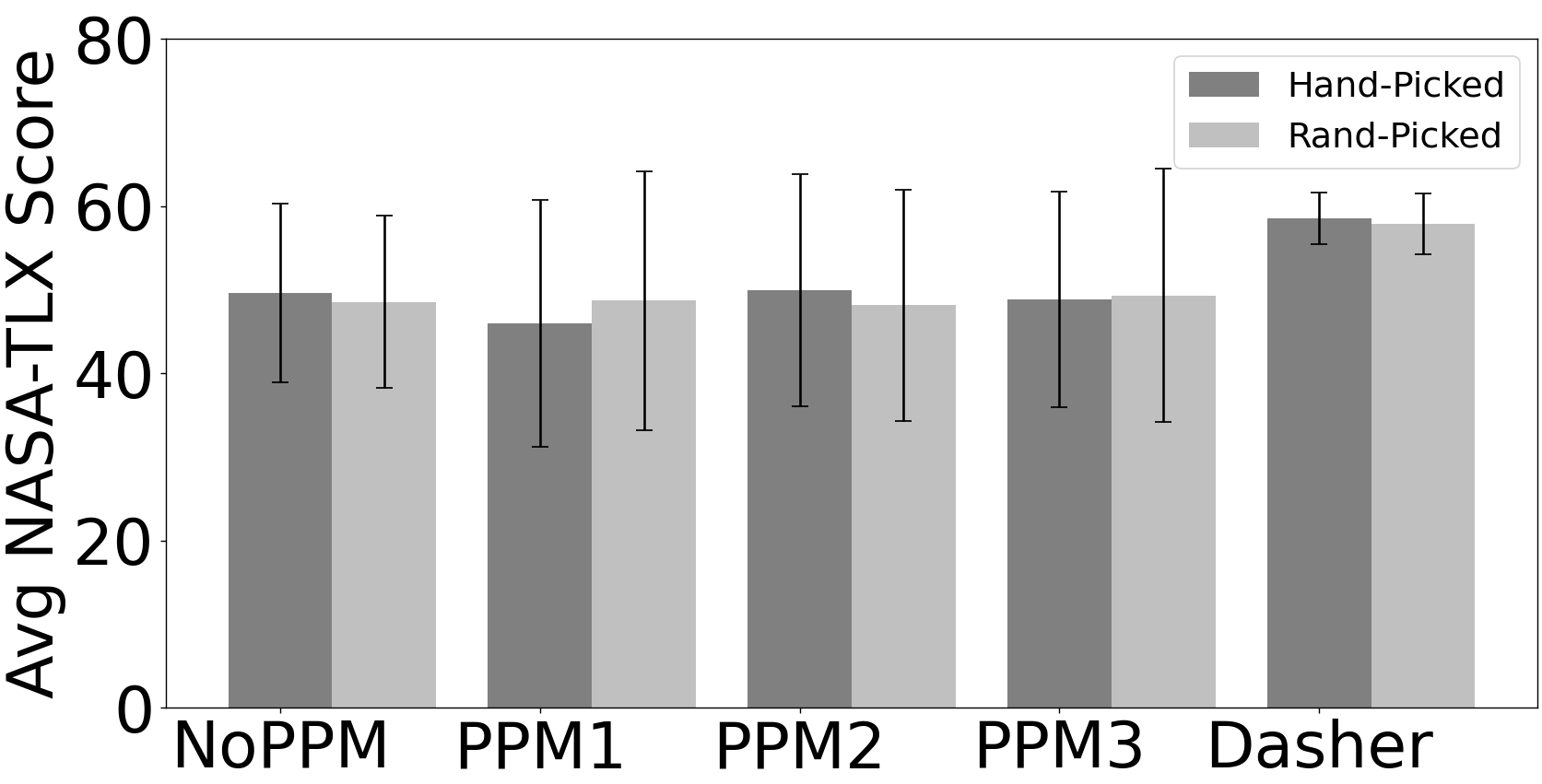}
         \caption{}
        \label{fig:sub1}
    \end{subfigure}
    \caption{Subjective evaluation outcomes: the average SUS score with a mouse (a) and an eye-tracker (b), the average NASA-TLX  score with a mouse (c) and an eye-tracker (d).}
    \label{fig 2: a and b}
\end{figure*}

The overall performance evaluation of the on-screen keyboards was undertaken based on the results collected from a typing experiment. The experimental conditions were divided into two categories. 


\subsection{Experiment 1}

Among all the results presented in Table~\ref{tab:my-table1}, for HandFT on Flex-Tree, the highest average speed was 30.37$\pm$9.18 letters/min with the mouse under the PPM2 condition and 16.05$\pm$2.68 letters/min with the eye-tracker under the PPM3 condition. For RandFT on Flex-Tree, the best average speed was 28.34$\pm$7.41 letters/min with the mouse and 16.33$\pm$2.97 letters/min with the eye-tracker, under the PPM2 and PPM3 conditions, respectively. The $ITR_{com}$ and $ITR_{letter}$ were 108.4$\pm$19.7 bits/min and 100.7$\pm$18.3 bits/min with the eye-tracker under PPM3 condition. Table~\ref{tab:tab3} indicates that for RandDS on Dasher, the highest average speed was 10.35$\pm$3.3032 letters/min with the mouse and 4.96$\pm$1.70 letters/min with the eye-tracker. For HandDS on Dasher, the best average speed was 9.18$\pm$3.02 letters/min with the mouse and 6.83$\pm$3.29 letters/min with the eye-tracker.


 \textit{1. PPM versus NoPPM with mouse:} For RandFT (see in Table~\ref{tab:my-table1}), the highest average speeds were 24.17$\pm$4.44 letters/min for NoPPM, 26.43$\pm$6.80 letters/min for PPM1, 28.34$\pm$7.41 letters/min for PPM2, and 27.76$\pm$8.95 letters/min for PPM3. For HandFT (see in Table~\ref{tab:my-table1}), the highest average speeds were 21.68$\pm$7.29 letters/min for NoPPM, 23.71$\pm$4.96 letters/min for PPM1, 30.37$\pm$9.18 letters/min for PPM2, and 27.55$\pm$7.73 letters/min for PPM3. 

 


\textit{2. PPM versus NoPPM with eye-tracker:} For RandFT (see in Table~\ref{tab:my-table1}), the highest average speeds were 13.84$\pm$2.10 letters/min for NoPPM, 15.24$\pm$1.88 letters/min for PPM1, 15.67$\pm$2.83 letters/min for PPM2, and 16.33$\pm$2.97 letters/min for PPM3. For HandFT (see in Table~\ref{tab:my-table1}), the best average speeds were 13.31$\pm$2.94 letters/min for NoPPM, 15.39$\pm$2.66 letters/min for PPM1, 15.76$\pm$3.41 letters/min for PPM2, and 16.05$\pm$2.68 letters/min for PPM3.

 
\textit{3. Dasher with a mouse versus an eye-tracker:} The highest average speed for Dasher was 10.35$\pm$3.30 letters/min in the RandDS condition with the mouse. With the eye-tracker, the best average speed was 6.83$\pm$3.29 letters/min in the HandDS condition (see Table~\ref{tab:tab3}).



\subsection{Experiment 2}

This experiment evaluates deletion times for Dasher and Flex-Tree with PPM3. Table~\ref{tab:tab3} shows that the fastest average deletion times were 1.42$\pm$0.19 seconds per letter for Flex-Tree and 4.26$\pm$0.19 seconds per letter for Dasher, both achieved with HandDS. For RandDS, the deletion times were 1.47$\pm$0.16 seconds per letter for Flex-Tree and 4.58$\pm$1.06 seconds per letter for Dasher. In both scenarios, Dasher had longer deletion times compared to Flex-Tree.


\section {Subjective Evaluation}

\subsection{System usability scale}

The system usability scale (SUS) is a widely used tool for evaluating subjective usability through ten Likert-type items, with scores ranging from 0 to 100. Higher scores reflect better usability, measuring effectiveness, efficiency, and satisfaction~\cite{brooke1996sus, meena2018toward, 9964264}. We administered the SUS for each experimental condition and calculated the average scores across participants for both mouse and eye-tracker inputs.

For HandFT with the mouse, the average SUS scores were 73.75$\pm$22.77 for NoPPM, 82.34$\pm$ 16.26 for PPM1, 83.12$\pm$ 16.41 for PPM2, and 79.21$\pm$17.43 for PPM3, with Dasher scoring 64.84$\pm$25.21. With the eye-tracker, the scores for HandFT were 77.81$\pm$18.90 for NoPPM, 82.50$\pm$16.93 for PPM1, 81.40$\pm$17.51 for PPM2, and 79.84$\pm$19.93 for PPM3, while Dasher scored 42.81$\pm$32.45. For RandFT with the mouse, the average SUS scores were 80.31$\pm$18.59 for NoPPM, 83.43$\pm$16.25 for PPM1, 83.90$\pm$14.94 for PPM2, and 82.50$\pm$19.19 for PPM3, with Dasher scoring 63.12$\pm$29.16. With the eye-tracker, the scores for RandFT were 76.25$\pm$20.95 for NoPPM, 82.96$\pm$16.48 for PPM1, 80.15$\pm$17.04 for PPM2, and 79.84$\pm$21.59 for PPM3, while Dasher scored 42.65$\pm$36.21. In both modalities, Flex-Tree with PPM1, PPM2, and PPM3 consistently outperformed both Flex-Tree with NoPPM (alphabetical layout)~\cite{meena2018toward} and Dasher, as shown in Fig.~\ref{fig 2: a and b} (a-b).

\subsection{NASA task load index}

The NASA Task Load Index (NASA-TLX) measures perceived workload across various tasks and systems, with scores ranging from 0 to 100, where lower scores signify better performance~\cite{raiha2012exploratory, meena2018toward, 9964264}. We used NASA-TLX to assess the workload associated with interacting with the on-screen keyboards, evaluating mental, physical, and temporal demands, as well as performance, effort, and frustration. The average scores for mouse and eye-tracker input modalities were calculated and shown in Fig.~\ref{fig 2: a and b} (c-d).

For the mouse input with HandFT, the average NASA-TLX scores were 41.25$\pm$17.60 for NoPPM, 38.95$\pm$20.60 for PPM1, 38.33$\pm$20.13 for PPM2, and 35.52$\pm$19.71 for PPM3, compared to 45.00$\pm$13.17 for Dasher. With the eye-tracker, the scores were 49.58$\pm$10.71 for NoPPM, 45.98$\pm$14.75 for PPM1, 49.94$\pm$13.92 for PPM2, and 48.80$\pm$12.89 for PPM3, while Dasher scored 58.54$\pm$3.11. For RandFT with the mouse, average NASA-TLX scores were 37.96$\pm$19.90 for NoPPM, 37.70$\pm$20.50 for PPM1, 37.81$\pm$19.92 for PPM2, and 39.47$\pm$18.10   for PPM3, with Dasher at 43.95$\pm$10.44. With the eye-tracker, the scores were  48.54$\pm$10.32 for NoPPM,  48.69$\pm$15.48 for PPM1, 48.12 $\pm$13.79 for PPM2, and 49.32$\pm$15.12 for PPM3, compared to 57.86$\pm$3.62 for Dasher. Overall, Flex-Tree with PPM1, PPM2, and PPM3 showed lower workload scores compared to Dasher in both modalities. Among these, PPM1, PPM2, and PPM3 consistently had lower scores than NoPPM with the eye-tracker, while scores for NoPPM, PPM1, PPM2, and PPM3 were relatively similar with the mouse.

\section{Discussion and Conclusion}

The proposed Flex-Tree is the first example of integrating the PPM technique with the tree-based on-screen keyboard using eye-tracking. Flex-Tree presents a viable alternative for beginner users who use eye-tracking. The typing spelling performance of Flex-Tree with PPM3 with a 1.5~s dwell time (16.33$\pm$2.97) surpassed the state-of-the-art results with 1.5~s dwell time methods~\cite{meena2016novel, meena2018toward, 9964264, 10.14236/ewic/HCI2018.148, cecotti2019multiscript, 7464272}. Further, Flex-Tree with PPM3 results are higher or approx than the adaptive time base methods in asynchronous mode~\cite{meena2019design}. However, further combining the Flex-Tree with the adaptive method could enhance typing spelling performance. For beginner users, the performance of Flex-Tree was higher than Dasher~\cite{tuisku2008now}. Flex-Tree received higher ratings on the system usability scale and lower ratings on the NASA Task Load Index for both input modalities than the Dasher and the alphabetical layout (NoPPM).

Our experiments identified PPM3 as the most effective, highlighting the importance of higher K values for enhancing Flex-Tree's performance, particularly with eye-tracking. Additionally, Flex-Tree outperformed Dasher in deletion times, establishing it as a superior alternative. As shown in Table ~\ref{tab:tab3}, Dasher's deletion times varied significantly, while Flex-Tree's deletion times were more consistent across characters.

We observed that typing randomly picked sentences was faster than hand-picked ones, indicating that our \textit{PredModel} significantly improves typing speed. Interestingly, PPM excelled at predicting the following letter within words but struggled after a space input. Introducing next-word prediction~\cite{10412159} could improve Flex-Tree further. Avoiding similar-looking characters like ``I," ``l," and ``\textbar{}" could also increase typing speed, as some users had difficulty distinguishing them. In eye-tracker settings, a concept worth exploring is displaying the next character to type instead of the ``Last\_five\_char\_field". Future research, including longitudinal studies, may offer deeper insights into Flex-Tree's performance over time. 

\subsection*{Acknowledgment}
This study was supported by the RES/MHRD/CS/P0294/
2324/0056-4224 and IP/IITGN/CSE/YM/2324/05 grants.

\bibliographystyle{unsrt}
\bibliography{mybib.bib}

\begin{thebibliography}{10}

\bibitem{WSOReprt}
World stroke organization (wso): Global stroke fact sheet 2022. accessed: Apr-14-2024 [online]. available:.
\newblock \url{https://www.world-stroke.org/assets/downloads/WSO_Global_Stroke_Fact_Sheet.pdf}.

\bibitem{howden2017sdg}
Philippa Howden-Chapman, Jose Siri, Elinor Chisholm, Ralph Chapman, Christopher~NH Doll, and Anthony Capon.
\newblock Sdg 3: Ensure healthy lives and promote wellbeing for all at all ages.
\newblock {\em A guide to SDG interactions: from science to implementation. Paris, France: International Council for Science}, pages 81--126, 2017.

\bibitem{ascari2021computer}
R~E de~Oliveira Schultz~Ascari, L~Silva, and R~Pereira.
\newblock Computer vision applied to improve interaction and communication of people with motor disabilities: A systematic mapping.
\newblock {\em Technology and Disability}, 33(1):11--28, Jan. 2021.

\bibitem{pinheiro2011alternative}
C~G Pinheiro, E~L Naves, P~Pino, E~Losson, A~O Andrade, and G~Bourhis.
\newblock Alternative communication systems for people with severe motor disabilities: a survey.
\newblock {\em BioMedical Engineering OnLine}, 10(1), Apr. 2011.

\bibitem{meena2018toward}
Yogesh~Kumar Meena, Hubert Cecotti, Kongfatt Wong-Lin, Ashish Dutta, and Girijesh Prasad.
\newblock Toward optimization of gaze-controlled human-computer interaction: Application to hindi virtual keyboard for stroke patients.
\newblock {\em IEEE Transactions on Neural Systems and Rehabilitation Engineering}, 26(4):911--922, 2018.

\bibitem{OSK}
Charlie Danger.
\newblock On-screen keyboards: Accessibility softwares. [online]. available:.
\newblock \url{https://www.bltt.org/software/osk.htm}.

\bibitem{ward2000dasher}
David~J Ward, Alan~F Blackwell, and David~JC MacKay.
\newblock Dasher—a data entry interface using continuous gestures and language models.
\newblock In {\em Proceedings of the 13th Annual ACM Symposium on User Interface Software and Technology}, pages 129--137, 2000.

\bibitem{hosni2019eeg}
S.M. Hosni, H.A. Shedeed, M.S. Mabrouk, et~al.
\newblock Eeg-eog based virtual keyboard: Toward hybrid brain computer interface.
\newblock {\em Neuroinformatics}, 17:323–341, 2019.

\bibitem{chakraborty2019eye}
Partha Chakraborty, Dipa Roy, Md.~Zahidur Rahman, and Saifur Rahman.
\newblock Eye gaze controlled virtual keyboard.
\newblock {\em International Journal of Recent Technology and Engineering (IJRTE)}, 8(4), November 2019.
\newblock Retrieval Number: D8049118419/2019©BEIESP.

\bibitem{cecotti2018multimodal}
Hubert Cecotti, Yogesh~Kumar Meena, and Girijesh Prasad.
\newblock A multimodal virtual keyboard using eye-tracking and hand gesture detection.
\newblock In {\em 2018 40th Annual International Conference of the IEEE Engineering in Medicine and Biology Society (EMBC)}, pages 3330--3333. IEEE, 2018.

\bibitem{meena2019design}
Yogesh~Kumar Meena, Hubert Cecotti, KongFatt Wong-Lin, and Girijesh Prasad.
\newblock Design and evaluation of a time adaptive multimodal virtual keyboard.
\newblock {\em Journal on Multimodal User Interfaces}, 13:343--361, 2019.

\bibitem{cecotti2019multiscript}
Hubert Cecotti, Yogesh~Kumar Meena, Braj Bhushan, Ashish Dutta, and Girijesh Prasad.
\newblock A multiscript gaze-based assistive virtual keyboard.
\newblock In {\em 2019 41st Annual International Conference of the IEEE Engineering in Medicine and Biology Society (EMBC)}, pages 1306--1309. IEEE, 2019.

\bibitem{tuisku2008now}
Outi Tuisku, P{\"a}ivi Majaranta, Poika Isokoski, and Kari-Jouko R{\"a}ih{\"a}.
\newblock Now dasher! dash away! longitudinal study of fast text entry by eye gaze.
\newblock In {\em Proceedings of the 2008 Symposium on Eye Tracking Research \& Applications}, pages 19--26, 2008.

\bibitem{drinic2003ppm}
Milenko Drinic, Darko Kirovski, and Miodrag Potkonjak.
\newblock Ppm model cleaning.
\newblock In {\em Data Compression Conference, 2003. Proceedings. DCC 2003}, pages 163--172. IEEE, 2003.

\bibitem{hu2010improving}
Yichuan Hu, Fazle Khan, and Yu~Li.
\newblock Improving ppm algorithm using dictionaries.
\newblock {\em arXiv preprint arXiv:1012.3790}, 2010.

\bibitem{10.14236/ewic/HCI2018.148}
Yogesh~Kumar Meena, Anirban Chowdhury, Ujjwal Sharma, Hubert Cecotti, Braj Bhushan, Ashish Dutta, and Girijesh Prasad.
\newblock A hindi virtual keyboard interface with multimodal feedback: a case study with a dyslexic child.
\newblock In {\em Proceedings of the 32nd International BCS Human Computer Interaction Conference}, HCI '18, Swindon, GBR, 2018. BCS Learning \& Development Ltd.

\bibitem{meena2016novel}
Yogesh~Kumar Meena, Hubert Cecotti, KongFatt Wong-Lin, and Girijesh Prasad.
\newblock A novel multimodal gaze-controlled hindi virtual keyboard for disabled users.
\newblock In {\em Proceedings of IEEE International Conference on Systems, Man, and Cybernetics}, pages 1--6, 2016.

\bibitem{9964264}
Yogesh~Kumar Meena, Hubert Cecotti, Braj Bhushan, Ashish Dutta, and Girijesh Prasad.
\newblock Detection of dyslexic children using machine learning and multimodal hindi language eye-gaze-assisted learning system.
\newblock {\em IEEE Transactions on Human-Machine Systems}, 53(1):122--131, 2023.

\bibitem{brooke1996sus}
John Brooke.
\newblock {\em Sus: a “quick and dirty” usability scale}.
\newblock Taylor and Francis, London, 1996.

\bibitem{raiha2012exploratory}
KJ~Raihä and S~Ovaska.
\newblock An exploratory study of eye typing fundamentals: Dwell time, text entry rate, errors, and workload.
\newblock In {\em Proceedings of the International ACM Conference on Human Factors in Computing Systems}, pages 3001--3010, 2012.

\bibitem{7464272}
Hubert Cecotti.
\newblock A multimodal gaze-controlled virtual keyboard.
\newblock {\em IEEE Transactions on Human-Machine Systems}, 46(4):601--606, 2016.

\bibitem{10412159}
Hisanori Fujii and Katsuma Ryo.
\newblock An implementation of word prediction for gaze swipe text entry.
\newblock In {\em 2023 Fourteenth International Conference on Mobile Computing and Ubiquitous Network (ICMU)}, pages 1--2, 2023.

\end{thebibliography}

\end{document}